\documentstyle[aps,psfig]{revtex}
\begin{document}
\draft
\twocolumn

\title{Probability Distributions and Particle Number Fluctuations of Trapped
Bose-Einstein Condensates in Different Dimensions}

\author{ Hongwei Xiong$^{1,2}$, Shujuan Liu$^{1}$, Guoxiang Huang$^{3}$, FengMin Wu$^{4}$, Zhijun Xu$^{1}$}

\address{$^{1}$Department of Applied Physics, Zhejiang
University of Technology, Hangzhou, 310032, China}
\address{$^{2}$Zhijiang College,
Zhejiang University of Technology, Hangzhou, 310012, China }
\address{$^{3}$Department of Physics, East China Normal University, Shanghai, 200062, 
China}
\address{$^{4}$ College of Mathematics and Physics, Zhejiang Normal University, 321004, China}

\date{\today}

\maketitle

\begin{abstract}
{\it The analytical probability distribution of finite systems
obeying Bose-Einstein statistics in one, two, and three dimensions
are investigated by using a canonical ensemble approach. Starting
from the canonical partition function of the system, a unified
approach is provided to study the probability distribution of a
condensate for various confinements and in different dimensions.
Based on the probability distribution function, it is
straightforward to obtain the mean ground state occupation number
and particle number fluctuations of the condensate. It is found
that the particle number fluctuations in a trapped Bose gas are
strongly dependent on the type of confining potential and on the
dimensionality of the system.}
\end{abstract}



\section{Introduction}


The experimental achievement of Bose-Einstein condensation (BEC) in trapped dilute
alkali atoms \cite{ALK}, spin-polarized hydrogen \cite{MIT} and metastable
helium \cite{HEL} has stimulated considerable theoretical research \cite
{RMP,PARKIN,LEG} on this unique phenomenon. The BEC has been recently realized
in quasi-one and quasi-two dimensions \cite{EXPLOW}, where new phenomena such as
quasicondensates with a fluctuating phase \cite{PETROV1,PETROV2,Kagan} and
a Tonks gas of impenetrable bosons \cite{PETROV2,Tonks,Olshanii} may be
observed. Among the several intriguing questions on the statistical
properties of the trapped Bose gas, the probability distribution and particle
number fluctuations $\left\langle \delta ^{2}N_{{\bf 0}}\right\rangle $ of
condensate play an important role. Apart from the intrinsic theoretical
interest, it is foreseeable that the condensate fluctuations will become
experimentally testable in the near future \cite{NEAR}. On the other hand,
the calculation of $\left\langle \delta ^{2}N_{{\bf 0}}\right\rangle $ is
important for investigating the phase collapse time of the condensate \cite
{JAV,WRI}. Recall that the BEC can not occur in
one-dimensional (1D) and two-dimensional (2D) uniform Bose gases at finite
temperature, because in that case thermal fluctuations can destabilize the condensate. 
The
realization of the trapped BECs in various dimensions makes the behavior of particle
number fluctuations a very interesting problem, especially for
1D and 2D harmonically trapped Bose gases.

Within a grand-canonical ensemble the particle number fluctuations of the
condensate are given by $\left\langle \delta ^{2}N_{{\bf {0}}}\right\rangle
=N_{{\bf {0}}}\left( N_{{\bf {0}}}+1\right) $, implying that $\delta N_{{\bf
{0}}}$ becomes of order $N$ when the temperature approaches zero. To avoid
this sort of unphysically large condensate fluctuations, a canonical (or a
microcanonical) ensemble has to be used to investigate $\delta N_{{\bf {0}}}$.
Because in the experiment the trapped atoms are cooled continuously from
the surrounding, the system can be taken as being in contact with a
heat bath but the total number of particles in the system is conserved. It
is therefore necessary to use the canonical ensemble to investigate the
statistical properties of the trapped Bose gas.

Within the canonical (or microcanonical) ensemble, the particle
number fluctuations have been studied  in the
case of three-dimensional (3D) ideal Bose gases confined in a box
\cite {HAUGE,FUJI,ZIF,BOR,WIL}, and in the presence of a harmonic
trap \cite {WIL,POL,GAJ,GRO,NAVEZ,BAL,HOL1,HOL2}. The question of
how interatomic interactions affect the particle number
fluctuations has also  been the object of several recent theoretical
investigations \cite {GIO,IDZ,ILLU,MEI,KOC,JAK,XIONG1}.
Although the phase fluctuations \cite{PETROV1,PETROV2,STOOF} of a
condensate have been studied for quasi-1D and quasi-2D Bose
gases, to best our knowledge up to now an analytical description
of the probability distribution and the behavior of $\delta
N_{{\bf {0}}}$ for trapped 1D and 2D Bose gases have not been given
directly from the microscopic statistics of the system. The
purpose of the present work is an attempt to calculate $\delta
N_{{\bf {0}}}$ of the  trapped ideal Bose gas in various dimensions,
using the analytical description of the probability distribution
obtained directly from the analysis of the canonical partition
function of the system. In addition, the probability distribution
of the condensate will be used to calculate the mean ground state
occupation number $\left\langle N_{{\bf {0}}}\right\rangle $.

The paper is organized as follows. Sec. II is devoted to outline the
canonical ensemble, which is developed to consider  the probability
distribution and particle number fluctuations of trapped ideal Bose gases.
In Secs. III-V  we investigate the particle number fluctuations of 3D, 2D and
1D Bose gases trapped in a harmonic potential and in a box, respectively.
The last section  (Sec. VI) contains a discussion and summary of our results.


\section{Probability Distribution and Particle Number Fluctuations of The
Condensate}


In the canonical ensemble, the canonical partition function of $N$ trapped
ideal bosons is given by

\begin{equation}
{Z}\left( N\right) {=\sum_{\Sigma _{{\bf {n}}}N_{{\bf {n}}}=N}\exp \left[
-\beta \Sigma _{{\bf {n}}}N_{{\bf {n}}}\varepsilon _{{\bf {n}}}\right] },
\label{par1}
\end{equation}

\noindent where $\beta=1/k_{B}T$,
 $N_{{\bf n}}$ and $\varepsilon _{{\bf n}}$ are occupation
number and energy level of the state ${\bf {n}}$, respectively.
The basic starting point of our approach  for calculating the probability 
distribution
of the condensate is   to separate off  the ground state labelled
${\bf {n}}={\bf 0}$ from the states ${\bf {n}}\neq {\bf 0}$:

\begin{equation}
{Z}\left( N\right) {=\sum_{N_{{\bf 0}}=0}^{N}\left\{ \exp \left[ -\beta N_{
{\bf 0}}\varepsilon _{{\bf 0}}\right] Z_{0}\left( N_{T}\right) \right\} },
\label{par2}
\end{equation}

\noindent where $Z_{0}\left( N_{T}\right) $ represents the canonical
partition function of a fictitious system consisting of $N_{T}=N-N_{{\bf 0}}$
trapped ideal non-condensed bosons:

\begin{equation}
{Z_{0}\left( N_{T}\right) =\sum_{\sum_{{\bf {n}}\neq {\bf 0}}N_{{\bf {n}}%
}=N_{T}}\exp \left[ -\beta \sum_{{\bf {n}}\neq {\bf 0}}N_{{\bf {n}}%
}\varepsilon _{{\bf {n}}}\right] .}  \label{II-function-1}
\end{equation}
\vspace{1pt}Note that the ground state ${\bf {n}}={\bf 0}$ has been
separated off in the canonical partition function ${Z_{0}\left( N_{T}\right)
}$ of the fictitious system. In the following we will find that this
separation plays a crucial role for the calculations of the probability
distribution of the system.

Assuming that $A_{0}\left( N,N_{{\bf 0}}\right) $ is the free energy of the
fictitious system, we have

\begin{equation}
{A_{0}(N,N_{{\bf 0}})=-k_{B}T\ln Z_{0}(N_{T}).}  \label{free-energy}
\end{equation}

The calculation of the free energy ${A_{0}\left( N,N_{{\bf 0}}\right) }$ is
nontrivial because there is a requirement that the number of non-condensed
bosons is $N_{T}$ in the summation of the canonical partition function $%
Z_{0}\left( N_{T}\right) $. To proceed we define a generating
function $G_0$ for $Z_{0}\left( N_{T}\right)$ in the following
manner. For any complex number $z$, let

\begin{equation}
{G_{0}\left( T,z\right) }={\sum_{N_{T}=0}^{\infty }}z{^{N_{T}}Z_{0}\left(
N_{T}\right) .}  \label{a-generate-define}
\end{equation}

\noindent The generating function can be evaluated easily. To obtain $%
Z_{0}\left( N_{T}\right) $ we note that by definition $Z_{0}\left(
N_{T}\right) $ is the coefficient of $z^{N_{T}}$ in the expansion of $%
G_{0}\left( T,z\right) $ in powers of $z$. Thus one has

\begin{equation}
{Z_{0}\left( N_{T}\right) }={\frac{1}{2\pi i}\oint d}z{\frac{G_{0}\left(
T,z\right) }{z^{N_{T}+1}},}  \label{a-relation}
\end{equation}

\noindent where the contour of integration is a closed path in the complex $%
z $ plane around $z=0$.

Assuming $\exp \left[ g\left( z\right) \right] =G_{0}\left(
T,z\right) /z^{N_{T}+1}$, the saddle point $z_{0}$ \cite{DAR} is determined
by $\partial g\left( z_{0}\right) /\partial z_{0}=0$. Expanding the
integrand of Eq. (\ref{a-relation}) about $z=z_{0}$, we have

\begin{equation}
{Z_{0}\left( N_{T}\right) }={\frac{1}{2\pi i}\oint d}z\exp \left[ g\left(
z_{0}\right) +o\left( (z-z_{0})^{2}\right) \right] {,}  \label{high-order}
\end{equation}

\noindent where $o\left( (z-z_{0})^{2}\right) =\frac{1}{2}\left(
z-z_{0}\right) ^{2}\frac{\partial ^{2}}{\partial z_{0}^{2}}g\left(
z_{0}\right) +\cdots $ represents the high order terms when expanding the
integrand about the saddle point $z_{0}$.

Omitting the high order terms on the right hand side of Eq. (\ref{high-order}%
), it is straightforward to obtain the following relations between the free
energy $A_{0}\left( N,N_{{\bf {0}}}\right) $ and the saddle point $z_{0}$ of
the fictitious $N-N_{{\bf 0}}$ non-condensed bosons

\begin{equation}
{-\beta \frac{\partial }{\partial N_{{\bf {0}}}}A_{0}\left( N,N_{{\bf {0}}%
}\right) =\ln z_{0}.}  \label{relation1}
\end{equation}

\noindent In addition, the saddle point $z_{0}$ is determined by

\begin{equation}
{N_{{\bf {0}}}=N-\sum_{{\bf {n}}\neq {\bf {0}}}\frac{1}{\exp \left[
\varepsilon _{{\bf {n}}}/k_{B}T\right] z_{0}^{-1}-1}.}  \label{relation2}
\end{equation}

\noindent Although the relations given by Eqs. (\ref{relation1}) and (\ref{relation2})
can not provide an explicit result of $Z_{0}\left( N_{T}\right) $, they are
very useful to calculate the probability distribution of the condensate.

We should stress that the relations given by Eqs. (\ref{relation1}) and 
(\ref{relation2})
are reliable although the disputable saddle-point method is used to
investigate the free energy of the fictitious system. It is well known that
the applicability of  the standard saddle-point method for condensed Bose
gases has been the subject of a long debate \cite{ZIF,DIN}. In
conventional approaches, the saddle-point method is used to discuss the
canonical partition function ${Z}\left( {N}\right) $. The generating
function is therefore defined by ${G\left( T,z\right) }={\sum_{N}^{\infty
}}z^{N}{Z}\left( {N}\right) ${. In this scheme,  the high order terms  of }$%
\ln Z\left( N\right) $ can not be omitted for the temperature below the
onset of the BEC, because the factor $1/\left( z_{0}^{-1}e^{\beta \epsilon _{%
{\bf {0}}}}-1\right) $ in the high order terms  would be on the order of $N$
(See Eq. (8) in Ref. \cite{HOL2}). In our approach, however, the
saddle-point method is used only to discuss the canonical partition function ${%
Z_{0}\left( N_{T}\right) }$ of the fictitious non-condensed bosons. Because
the ground state has been separated off in ${Z_{0}\left( N_{T}\right) }$,
the high order terms of $\ln {Z_{0}\left( N_{T}\right) }$ can be safely
omitted.

Using the free energy of the fictitious system, the canonical
partition function of the system becomes

\begin{equation}
{Z}\left( {N}\right) {=\sum_{N_{{\bf 0}}=0}^{N}\exp \left[ q\left( N,N_{{\bf %
0}}\right) \right] },  \label{par3}
\end{equation}

\noindent where

\begin{equation}
{q\left( N,N_{{\bf 0}}\right) =-\beta N_{{\bf 0}}\varepsilon _{{\bf 0}%
}-\beta A_{0}\left( N,N_{{\bf 0}}\right) .}  \label{q-function}
\end{equation}

\noindent Obviously,  $P\left( N_{{\bf {0}}}|N\right) =\exp \left[ q\left(
N,N_{{\bf {0}}}\right) \right] /Z\left( N\right) $ represents the
probability to find $N_{{\bf 0}}$ atoms in the condensate.

Once $q\left( N,N_{{\bf 0}}\right) $ is calculated from the canonical
partition function of the system, the statistical properties of the system
can be easily obtained. However, it seems to be very difficult  to obtain
the analytical result of $q\left( N,N_{{\bf 0}}\right) $ directly from the
canonical partition function. To avoid this difficulty, we turn to
investigate the partial derivative of $q\left( N,N_{{\bf 0}}\right) $ with respect 
to $N_{{\bf 0}}$. Assuming

\begin{equation}
{\frac{\partial }{\partial N_{{\bf 0}}}q\left( N,N_{{\bf 0}}\right) =\alpha
\left( N,N_{{\bf 0}}\right) ,}  \label{q-alpha}
\end{equation}

\noindent from Eq. (\ref{q-function}) one obtains

\begin{equation}
{-\beta \frac{\partial }{\partial N_{{\bf 0}}}A_{0}(N,N_{{\bf 0}})=\beta {%
\varepsilon }_{{\bf 0}}+\alpha \left( N,N_{{\bf {0}}}\right).}
\end{equation}

\noindent Using the relations given by Eqs. (\ref{relation1}) and (\ref
{relation2}), we get

\begin{equation}
{N_{{\bf 0}}=N-\sum_{{\bf n\neq 0}}\frac{1}{\exp \left[ \beta \left(
\varepsilon _{{\bf {n}}}-{{\varepsilon }_{{\bf 0}}}\right) \right] \exp
\lbrack -{\alpha \left( N,N_{{\bf {0}}}\right) }\rbrack -1}.}  \label{main3}
\end{equation}

The most probability to find $N_{{\bf 0}}$ atoms in the condensate
is determined by  requiring  $\frac{\partial}{\partial N_{{\bf
0}}} q\left( N,N_{{\bf 0}}\right) =0$. Thus the most probable
value ${N_{{\bf 0} }^{p}}$ is determined by setting ${\alpha
\left( N,N_{{\bf {0}}}\right) =0}$ in the right hand side of Eq.
(\ref{main3})

\begin{equation}
{N_{{\bf {0}}}^{p}=N-\sum_{{\bf {n}}\neq {\bf 0}}\frac{1}{\exp
\left[ \beta \left( \varepsilon _{{\bf {n}}}-{{\varepsilon }_{{\bf
0}}}\right) \right] -1} .}  \label{non1}
\end{equation}

\noindent It is interesting to note that $N_{{\bf {0}}}^{p}$ is exactly the
mean ground state occupation number in the frame of a grand-canonical
ensemble. For sufficiently large $N$, the sum $\sum_{N_{{\bf {0}}}=0}^{N}$
in Eq. (\ref{par3}) may be replaced by the largest term, since  the error
omitted in doing so will be statistically negligible. Hence  the result given
by Eq. (\ref{non1}) shows the equivalence between the canonical and
grand-canonical ensembles for large $N$.

From the results given by Eqs. (\ref{main3}) and (\ref{non1}), we get
the following equation for determining $\alpha \left( N,N_{{\bf {0}}}\right)
$

$${
N_{{\bf {0}}}\vspace{1pt}-N_{{\bf {0}}}^{p}=\sum_{{\bf {n}}\neq {\bf {0}}}
\frac{1}{\exp \left[ \beta \left( \varepsilon _{{\bf {n}}}-{{\varepsilon }_{
{\bf 0}}}\right) \right] -1}
}$$

\begin{equation}
{ ~~~~~~ -\sum_{{\bf {n}}\neq {\bf {0}}}\frac{1}{\exp
\left[ \beta \left( \varepsilon _{{\bf {n}}}-{{\varepsilon }_{{\bf 0}}}
\right) \right] \exp \lbrack -{\alpha \left( N,N_{{\bf {0}}}\right) }\rbrack
-1}.}  \label{alpha}
\end{equation}

\noindent Thus once we know the single-particle energy level of the system, it is
straightforward to obtain $\alpha \left( N,N_{{\bf {0}}}\right) $. Using $%
\alpha \left( N,N_{{\bf {0}}}\right) $, one can obtain the probability
distribution of the system.

From Eq. (\ref{q-alpha}), we get the following result for $q\left( N,N_{%
{\bf {0}}}\right) $

\begin{equation}
{q\left( N,N_{{\bf {0}}}\right) =\int_{N_{{\bf {0}}}^{p}}^{N_{{\bf {0}}%
}}\alpha \left( N,N_{{\bf {0}}}\right) dN_{{\bf {0}}}+q\left( N,N_{{\bf {0}}%
}^{p}\right) .}  \label{qalpha2}
\end{equation}

\noindent The partition function of the system is then

\begin{equation}
{Z}\left( N\right) {=\sum_{N_{{\bf {0}}}=0}^{N}\left\{ \exp \left[ q\left(
N,N_{{\bf {0}}}^{p}\right) \right] G\left( N,N_{{\bf {0}}}\right) \right\} ,}
\label{par-alpha}
\end{equation}

\noindent where

\begin{equation}
{G\left( N,N_{{\bf {0}}}\right) =\exp \left[ \int_{N_{{\bf {0}}}^{p}}^{N_{%
{\bf {0}}}}\alpha \left( N,N_{{\bf {0}}}\right) dN_{{\bf {0}}}\right] .}
\label{ideal-dis}
\end{equation}

\noindent It is obvious that ${G\left( N,N_{{\bf {0}}}\right) }$ represents the ratio $P\left(
N_{{\bf {0}}}|N\right) /P(N_{{\bf {0}}}^{p}|N)$, {\it i.e.,} the relative
probability to find $N_{{\bf 0}}$ atoms in the condensate. From Eq. (\ref
{ideal-dis}), the normalized probability distribution function is given by

\begin{equation}
{G}_{n}{\left( N,N_{{\bf {0}}}\right) =A}_{n}{\exp \left[ \int_{N_{{\bf {0}}%
}^{p}}^{N_{{\bf {0}}}}\alpha \left( N,N_{{\bf {0}}}\right) dN_{{\bf {0}}%
}\right] ,}  \label{norm-dis}
\end{equation}

\noindent where $A_{n}$ is a normalization constant.

As soon as we know $G\left( N,N_{{\bf {0}}}\right) $, the statistical
properties of the system can be clearly described. From Eqs. (\ref{par-alpha}%
) and (\ref{ideal-dis}) one obtains $\left\langle N_{{\bf {0}}}\right\rangle
$ and $\left\langle \delta ^{2}N_{{\bf {0}}}\right\rangle $ within the
canonical ensemble:

\begin{equation}
{\left\langle N_{{\bf {0}}}\right\rangle = \frac{ \sum_{N_{{\bf {0}}%
}=0}^{N}N_{{\bf {0}}}\exp \left[ q\left( N,N_{{\bf {0}}}\right) \right] } {%
\sum_{N_{{\bf {0}}}=0}^{N}\exp \left[ q\left( N,N_{{\bf {0}}}\right) \right]}%
= \frac{\sum_{N_{{\bf {0}}}=0}^{N}N_{{\bf {0}}}G\left( N,N_{{\bf {0}}%
}\right) } {\sum_{N_{{\bf {0}}}=0}^{N}G\left( N,N_{{\bf {0}}}\right) },}
\label{mean-ideal}
\end{equation}

$${
\left\langle \delta ^{2}N_{{\bf {0}}}\right\rangle =\left\langle N_{{\bf {0}
}}^{2}\right\rangle -\left\langle N_{{\bf {0}}}\right\rangle ^{2}
}$$

\begin{equation}
{ =\frac{
\sum_{N_{{\bf {0}}}=0}^{N}N_{{\bf {0}}}^{2}G\left( N,N_{{\bf {0}}}\right) }{
\sum_{N_{{\bf {0}}}=0}^{N}G\left( N,N_{{\bf {0}}}\right) }-\left[ \frac{
\sum_{N_{{\bf {0}}}=0}^{N}N_{{\bf {0}}}G\left( N,N_{{\bf {0}}}\right) }{
\sum_{N_{{\bf {0}}}=0}^{N}G\left( N,N_{{\bf {0}}}\right) }\right] ^{2}.}
\label{fluc-ideal}
\end{equation}

\noindent We see that in our approach, the calculation of
$\alpha \left( N,N_{{\bf {0}}}\right)
$ by using Eq. (\ref{alpha}) plays a crucial role to discuss the particle
number fluctuations of the condensate. The probability distribution of the
condensate is obtained from $\alpha \left( N,N_{{\bf {0}}}\right) $. We can
give a fairly accurate description of the particle number fluctuations through the
calculations of the probability distribution, although the high order terms
are omitted when obtaining  the relations  (\ref{relation1}) and (\ref
{relation2}). Our discussions on the
particle number fluctuations are reasonable mainly due to  the following
reasons:

(i) Because the ground state (${\bf n}={\bf 0}$) and the excited
states (${\bf n}\neq {\bf 0}$) have been separated off when
considering the partitioin function of the system, and the
saddle-point approximation is only used to investigate the
canonical partition function of the fictitious non-condensed
bosons, ${Z_{0}\left( N_{T}\right) }$, the high order terms
omitted in the approximation do not give a significant correction
to the particle number fluctuations.

(ii) When obtaining $\alpha \left( N,N_{{\bf {0}}}\right) $ through the
difference between $N_{{\bf {0}}}$ and $N_{{\bf {0}}}^{p}$, the high order
terms omitted in Eqs. (\ref{main3}) and (\ref{non1}) are nearly cancelled with
each other. This is true especially for the case of $\delta N_{{\bf {0}}}/N<<1$. Thus 
the
error coming from the high order terms will be lowered further for the
calculation of  $\alpha \left( N,N_{{\bf {0}}}\right) $.

On the other hand, in usual statistical method, ${\left\langle N_{{\bf 
{0}}}\right\rangle }$
and ${\left\langle \delta ^{2}N_{{\bf {0}}}\right\rangle }$ are obtained
through the first and second partial derivative of the canonical partition
function, respectively. When the saddle-point approximation is used to
calculate the canonical partition function of the system, the contribution
due to the high order terms are amplified in the second partial
derivative of the canonical partition function. Thus one can not obtain
accurate particle number fluctuations  using this method. The approach developed 
above provides in some sense a simple method recovering the
applicability of the saddle-point method through the calculations of the
probability distribution of the system.


\section{Three-Dimensional Bose Gases}


Now we apply the formulation presented in the last section to calculate the
thermodynamical quantities such as the particle
number fluctuations of the condensate for trapped ideal Bose gases.
An important feature characterizing the available magnetic trap is that the
confining potential can be safely approximated with the quadratic form

\begin{equation}
V_{ext}\left( {\bf r}\right) =\frac{m}{2}\left( \omega _{x}^{2}x^{2}+\omega
_{y}^{2}y^{2}+\omega _{z}^{2}z^{2}\right) ,  \label{potential}
\end{equation}

\noindent where $m$ is the mass of atoms. $\omega _{x}$, $\omega _{y}$, $\omega _{z}$
are frequencies of magnetic traps. The single-particle energy level of a 3D
harmonically confined ideal Bose gase has the form

\begin{equation}
{\varepsilon _{{\bf {n}}}=\left( n_{x}+\frac{1}{2}\right) \hbar \omega
_{x}+\left( n_{y}+\frac{1}{2}\right) \hbar \omega _{y}+\left( n_{z}+\frac{1}{%
2}\right) \hbar \omega _{z},}  \label{energy-trap}
\end{equation}

\noindent where $n_x, n_y$ and $n_z$ are non-negative integers.

Using the following density of states \cite{FIN,LIU}

\begin{equation}
{\rho \left( E\right) =\frac{1}{2}\frac{E^{2}}{\left( \hbar \omega
_{3D}\right) ^{3}}+\frac{3\overline{\omega }E}{2\omega _{3D}\left( \hbar
\omega _{3D}\right) ^{2}},}  \label{density-state}
\end{equation}

\noindent one obtains $N_{{\bf {0}}}$ from Eq. (\ref{main3})

$${
N_{{\bf {0}}}=N-}\frac{N}{\zeta \left( 3\right) }{\left( \frac{T}{T_{3D}}%
\right) ^{3}g_{3}\left[ \exp \left( \alpha \left( N,N_{{\bf {0}}}\right)
\right) \right] 
}$$

\begin{equation}
{ ~~~~~~~~~~~- \frac{3\overline{\omega }\zeta \left( 2\right) }{2\omega
_{3D}\left[ \zeta \left( 3\right) \right] ^{2/3}}\left( \frac{T}{T_{3D}}%
\right) ^{2}N^{2/3},}  \label{ideal3D}
\end{equation}

\noindent where $T_{3D}=\frac{\hbar \omega _{3D}}{k_{B}}\left( \frac{N}{
\zeta \left( 3\right) }\right) ^{1/3}$ is the critical temperature of the 3D
ideal Bose gas. $\overline{\omega }=\left(
\omega _{x}+\omega _{y}+\omega _{z}\right) /3$ and $\omega _{3D}=\left(
\omega _{x}\omega _{y}\omega _{z}\right) ^{1/3}$ are arithmetic and
geometric averages of the oscillator frequencies, respectively. $g_{3}\left(
z\right) $ belongs to the class of functions $g_{\alpha }\left( z\right)
=\sum_{n=1}^{\infty }z^{n}/n^{\alpha }$ and $\zeta \left( n\right) $ is
Riemann $\zeta $ function. 
Setting $\alpha \left( N,N_{{\bf {0}}}\right) =0$ in Eq. (\ref{ideal3D}),
the most probable value ${N_{{\bf {0}}}^{p}}$ is then

\begin{equation}
{N_{{\bf {0}}}^{p}=N-N\left( \frac{T}{T_{3D}}\right) ^{3}-\frac{3\overline{%
\omega }\zeta \left( 2\right) }{2\omega _{3D}\left[ \zeta \left( 3\right)
\right] ^{2/3}}\left( \frac{T}{T_{3D}}\right) ^{2}N^{2/3}.}  \label{most3D}
\end{equation}

\noindent From Eqs. (\ref{ideal3D}) and (\ref{most3D}) we obtain the result for $%
\alpha \left( N,N_{{\bf {0}}}\right) $

\begin{equation}
{\alpha \left( N,N_{{\bf 0}}\right) =-\frac{\zeta \left( 3\right) \left( N_{%
{\bf {0}}}-N_{{\bf {0}}}^{p}\right) }{\zeta \left( 2\right) Nt^{3}},}
\label{alpha-har}
\end{equation}

\noindent where $t=T/T_{3D}$
is a reduced temperature. When
obtaining $\alpha \left( N,N_{{\bf {0}}}\right) $ we have used the expansion
$g_{3}\left( 1+\delta \right) \approx \zeta \left( 3\right) +\zeta \left(
2\right) \delta $ \cite{ROB}. From Eqs. (\ref{ideal-dis}) and (\ref
{alpha-har}) we obtain the normalized probability distribution of the 3D
harmonically trapped ideal Bose gas:

\begin{equation}
{G_{3D}\left( N,N_{{\bf 0}}\right) =A}_{3D}{\exp \left[ -\frac{\zeta \left(
3\right) \left( N_{{\bf 0}}-N_{{\bf 0}}^{p}\right) ^{2}}{2\zeta \left(
2\right) Nt^{3}}\right] ,}  \label{3D-dis}
\end{equation}

\noindent where $A_{3D}$ is a normalization constant. It is interesting to
note that Eq. (\ref{3D-dis}) is a Gaussian distribution function.

From the formulas (\ref{mean-ideal}), (\ref{fluc-ideal}), (\ref{most3D}),
and (\ref{3D-dis}) we can obtain $\left\langle N_{{\bf {0}}}\right\rangle $
and $\left\langle \delta ^{2}N_{{\bf {0}}}\right\rangle $ of the system.
Shown in Fig. 1 is the the numerical result of $\delta N_{{\bf {0%
}}}=\sqrt{\left\langle \delta ^{2}N_{{\bf {0}}}\right\rangle }$ (solid line)
for $N=10^{3}$ ideal bosons in an isotropic harmonic trap. The dashed line
displays the result of Holthaus $et$ $al.$ \cite{HOL2}, where the
saddle-point method is developed to discuss the high order terms omitted in
the standard saddle-point approximation below the onset of BEC.

\begin{figure}[tb]
\psfig{figure=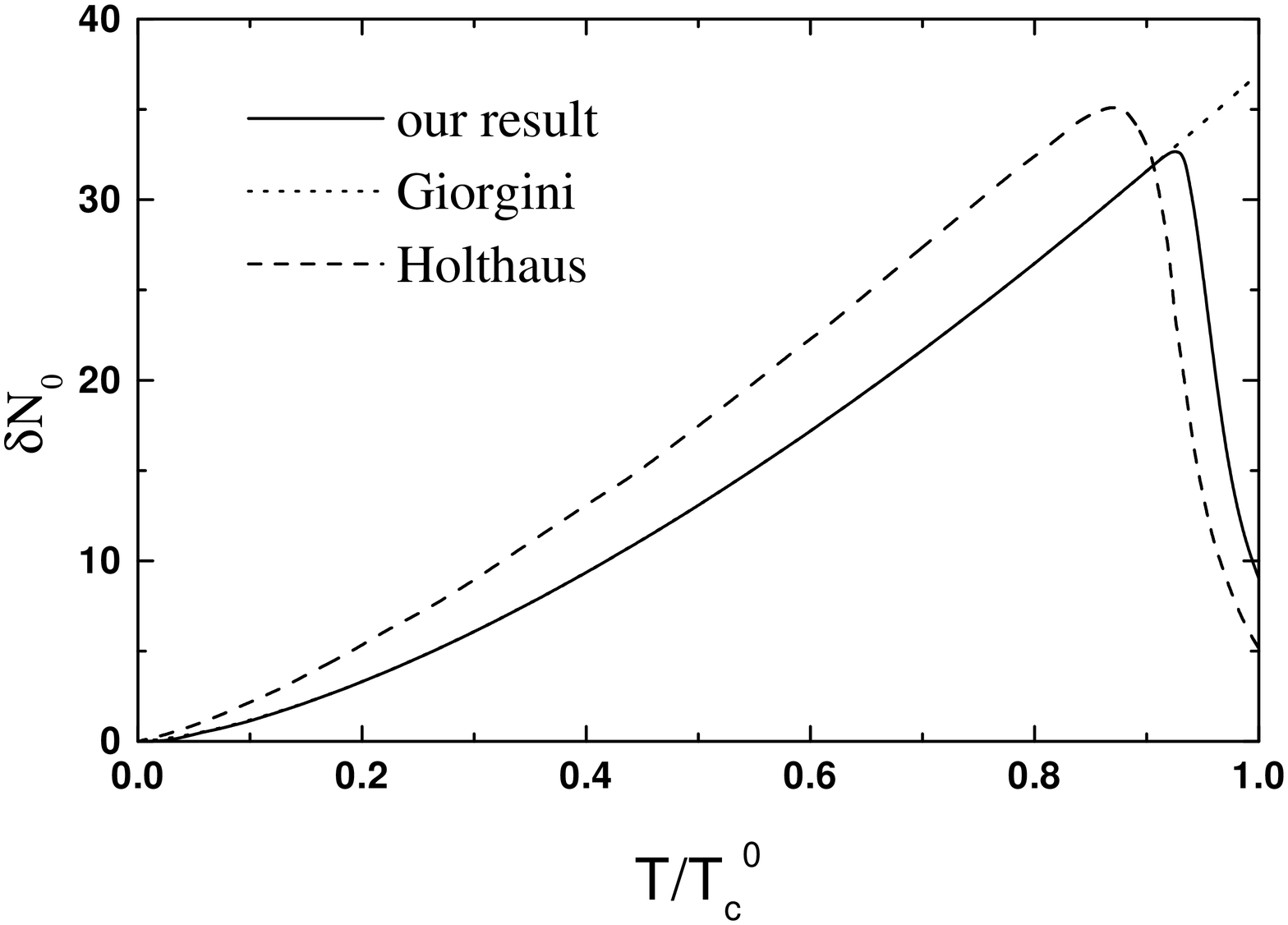,width=\columnwidth,angle=270}
\caption{Root-mean-square fluctuations $\delta N_{{\bf {0}}}$ for $N=10^{3}$
non-interacting bosons confined in a 3D isotropic harmonic trap. The solid
line displays $\delta N_{{\bf {0}}}$ obtained from the probability
distribution function Eq. (\ref{3D-dis}), while the dashed line shows the
result of Holthaus {\it et al. }\lbrack 27\rbrack . The dotted line displays
the result of Giorgini {\it et al.} \lbrack 28\rbrack\ (See Eq. (\ref
{analytical})). Near the critical temperature, our result agrees with that
of Holthaus {\it et al.} \lbrack 27\rbrack .}
\end{figure}

In Fig. 1 the dotted line displays the result of \cite{POL,GIO}. Our results
coincide with that of \cite{POL,GIO}, except near the critical temperature.
In fact, below the critical temperature, from the probability distribution
 (\ref{3D-dis}), one obtains the analytical result of the condensate
fluctuations

\begin{equation}
{\left\langle \delta ^{2}N_{{\bf 0}}\right\rangle =\frac{\zeta \left(
2\right) }{\zeta \left( 3\right) }Nt^{3}.}  \label{analytical}
\end{equation}

\noindent When obtaining Eq. (\ref{analytical}) we have used Eq. (\ref{a7})
derived in the Appendix. Below the critical temperature, Eq. (\ref
{analytical}) recovers the result in \cite{POL,NAVEZ,GIO} by noting that $%
\zeta \left( 2\right) =\pi ^{2}/6$. This shows the validity of the
probability distribution given by Eq. (\ref{3D-dis}) for discussing  the
statistical properties of the system. At the critical temperature $T_{3D}$,
however, our results give

\begin{equation}
{\left\langle \delta ^{2}N_{{\bf 0}}\right\rangle |_{T=T_{3D}}=\frac{\left(
1-2/\pi \right) \zeta \left( 2\right) N}{\zeta \left( 3\right) },}
\label{critical-fluc}
\end{equation}

\noindent which is much smaller than that obtained in Ref. \cite{GIO}. This
difference is apprehensible because the analysis in Ref.
\cite{GIO} holds in the canonical ensemble except near and above $T_{3D}$,
while our result holds also for the temperature near $T_{3D}$. Near the
critical temperature, our result (solid line) agrees with that of Holthaus
{\it et al.} \cite{HOL2}. Results given by Eqs. (\ref{analytical}) and (\ref
{critical-fluc}) show that particle number fluctuations in the 3D harmonic trap
have a normal behavior, ${\it i.e.}$, they are proportional to $N$.

When $T\rightarrow 0$, the particle number fluctuations of the condensate
should be zero. To check this, note that as $T\rightarrow 0$, ${G_{3D}\left(
N,N_{{\bf {0}}}\right) =A}_{3D}$ if $N_{{\bf {0}}}=N$, while ${G_{3D}\left(
N,N_{{\bf {0}}}\right) \rightarrow 0}$ when $N_{{\bf {0}}}\neq N$.
Additionally, $N_{{\bf {0}}}^{p}\rightarrow N$ in the case of $T\rightarrow 0
$. Thus, we obtain $\left\langle N_{{\bf {0}}}\right\rangle \rightarrow N$
and $\left\langle \delta ^{2}N_{{\bf {0}}}\right\rangle \rightarrow 0$ when $%
T\rightarrow 0$. The correct description of $\delta N_{{\bf {0}}}$ near zero
termperature and critical temperature again shows the validity of our method.

It is well known that confinement reduces the fluctuation effect. Thus
the fluctuations for bosons confined in a box should have a stronger dependence
on $N$, in comparison with the case confined in the harmonic trap.
To show this we consider a 3D ideal Bose gas confined in a cubic box. The
single-particle energy level in this case is given by $\varepsilon _{{\bf {n}}}=\pi
^{2}\hbar ^{2}(n_{x}^{2}+n_{y}^{2}+n_{z}^{2})/2mL^{2}$, where $L$ is the length of
the box. From Eqs. (\ref
{alpha}) and (\ref{ideal-dis}), the distribution function of the condensate
is given by

\begin{equation}
{G_{3B}\left( N,N_{{\bf {0}}}\right) =A}_{3B}{\exp \left[ -\frac{\lambda ^{6}%
}{12\pi V^{2}}\left| N_{{\bf {0}}}-N_{{\bf {0}}}^{p}\right| ^{3}\right] ,}
\label{3DBox}
\end{equation}

\noindent where $V=L^3$,  $A_{3B}$ and $\lambda =\sqrt{2\pi \beta \hbar ^{2}/m}$ are
normalization constant and thermal wavelength, respectively. From Eq. (\ref
{3DBox}) we obtain the particle number fluctuations below the critical
temperature:

\begin{equation}
{\left\langle \delta ^{2}N_{{\bf 0}}\right\rangle =A}\left( \frac{mk_{B}T}{%
\hbar ^{2}}\right) ^{2}V^{4/3}{,}  \label{3D-flu-box}
\end{equation}

\noindent where the coefficient $A=2/\pi ^{4}\times \Sigma _{{\bf {n}}\neq {\bf 0}}1/%
{\bf {n}}^{4}=0.105$. We see that, different from the harmonic trap, the particle 
number
fluctuations of the 3D Bose gas in a box exhibit an anomalous behavior, ${\it i.e.
}$,  it is  proportional to $V^{4/3}$ (or $N^{4/3}$).


\section{Two-Dimensional Bose Gases}


In this section we turn to discuss trapped 2D Bose gases.
The single-particle energy level of a 2D harmonically confined ideal Bose gas
takes the form

\begin{equation}
{\varepsilon _{{\bf {n}}}=\left( n_{x}+\frac{1}{2}\right) \hbar \omega
_{x}+\left( n_{y}+\frac{1}{2}\right) \hbar \omega _{y}.}  \label{2D-energy}
\end{equation}

\noindent From Eq. (\ref{main3}) one has

$${
N_{{\bf 0}}=N- 
}$$

\begin{equation}
{ \sum_{{\bf n}\neq {\bf 0}}\frac{1}{\exp \left[ \beta \left(
n_{x}\hbar \omega _{x}+n_{y}\hbar \omega _{y}\right) \right] \exp \lbrack -{%
\alpha \left( N,N_{{\bf {0}}}\right) }\rbrack -1}.}  \label{2D-1}
\end{equation}

\noindent Then we obtain ${N_{{\bf 0}}}$ through a simple 
integration over ${\bf n}$ \cite{MULLIN}:

\begin{equation}
{N_{{\bf 0}}=N-}\left( \frac{k_{B}T}{\hbar \omega _{2D}}\right)
^{2}g_{2}\left( \alpha \right) -\frac{\lbrack \zeta (2)\rbrack
^{1/2}k_{B}T\ln N}{\hbar \omega _{2D}}{,}  \label{2D-N0}
\end{equation}

\noindent where $\omega _{2D}=(\omega _{x}\omega _{y})^{1/2}$. 
When $0<\alpha <<1$, there is a good
approximation \cite{ROB} for $g_{2}\left( \alpha \right)$:

\begin{equation}
g_{2}\left( \alpha \right) \approx \zeta \left( 2\right) +\alpha \left(
1-\ln \alpha \right) {.}  \label{alpha1}
\end{equation}
In the case of $\alpha <0$ and $\left| \alpha \right| <<1$, $g_{2}\left( \alpha \right) 
$ is

\begin{equation}
g_{2}\left( \alpha \right) \approx \zeta \left( 2\right) -\left| \alpha
\right| \left( 1-\ln \left| \alpha \right| \right) {.}  \label{alpha2}
\end{equation}
By setting $\alpha =0$ in Eq. (\ref{2D-N0}), we obtain the most probable value ${N_{%
{\bf 0}}^{p}}$ as:

\begin{equation}
{N_{{\bf 0}}^{p}=N-}\left( \frac{k_{B}T}{\hbar \omega _{2D}}\right)
^{2}\zeta \left( 2\right) -\frac{\lbrack \zeta (2)\rbrack ^{1/2}k_{B}T\ln N}{%
\hbar \omega _{2D}}{.}  \label{2D-Np}
\end{equation}

Using the approximations given by Eqs. (\ref{alpha1}) and (\ref{alpha2}),
we arrive at the following result for ${\alpha \left( N,N_{{\bf {0}}}\right) }
$:

\begin{equation}
{\alpha \left( N,N_{{\bf {0}}}\right) =}-\frac{\zeta \left( 2\right) \zeta
\left( 3\right) \left( N_{{\bf {0}}}-{N_{{\bf 0}}^{p}}\right) }{2Nt^{2}\ln
\left( Nt^{2}/\zeta \left( 2\right) \right) }{,}  \label{2D-alpha}
\end{equation}

\noindent where $T_{2D}=\left( \frac{N}{\zeta \left( 2\right) }\right) ^{1/2}\frac{%
\hbar \omega _{2D}}{k_{B}}$ is the critical temperature of the
system in the thermodynamic limit.
Accordingly,  the  probability distribution of the system reads

\begin{equation}
G_{2D}{\left( N,N_{{\bf {0}}}\right) =A}_{2D}{\exp }\left[ -\frac{\zeta
\left( 2\right) \zeta \left( 3\right) \left( N_{{\bf {0}}}-{N_{{\bf 0}}^{p}}%
\right) ^{2}}{4Nt^{2}\ln \left( Nt^{2}/\zeta \left( 2\right) \right) }%
\right] {,}  \label{2D-dis}
\end{equation}

\noindent where $A_{2D}$ is a normalization constant. We see that  the probability
distribution of the 2D harmonically trapped ideal Bose gas is also
a Gaussian function. However, the behavior of the fluctuations is
different from that of the trapped 3D Bose gas because of the
factor $\ln \left( Nt^{2}/\zeta \left( 2\right) \right) $ in Eq.
(\ref{2D-dis}). Below the critical temperature, the analytical
description of the particle number fluctuations is given by

\begin{equation}
{\left\langle \delta ^{2}N_{{\bf 0}}\right\rangle =\frac{2Nt^{2}\ln \left(
Nt^{2}/\zeta \left( 2\right) \right) }{\zeta \left( 2\right) \zeta \left(
3\right) }.}  \label{2D-flu}
\end{equation}

\noindent At the critical temperature  $T_{2D}$, the particle number fluctuations takes the 
form

\begin{equation}
{\left\langle \delta ^{2}N_{{\bf 0}}\right\rangle |}_{T=T_{2D}}{=\frac{%
2\left( 1-2/\pi \right) N\ln \left( N/\zeta \left( 2\right) \right) }{\zeta
\left( 2\right) \zeta \left( 3\right) }.}  \label{2D-near}
\end{equation}

\begin{figure}[tb]
\psfig{figure=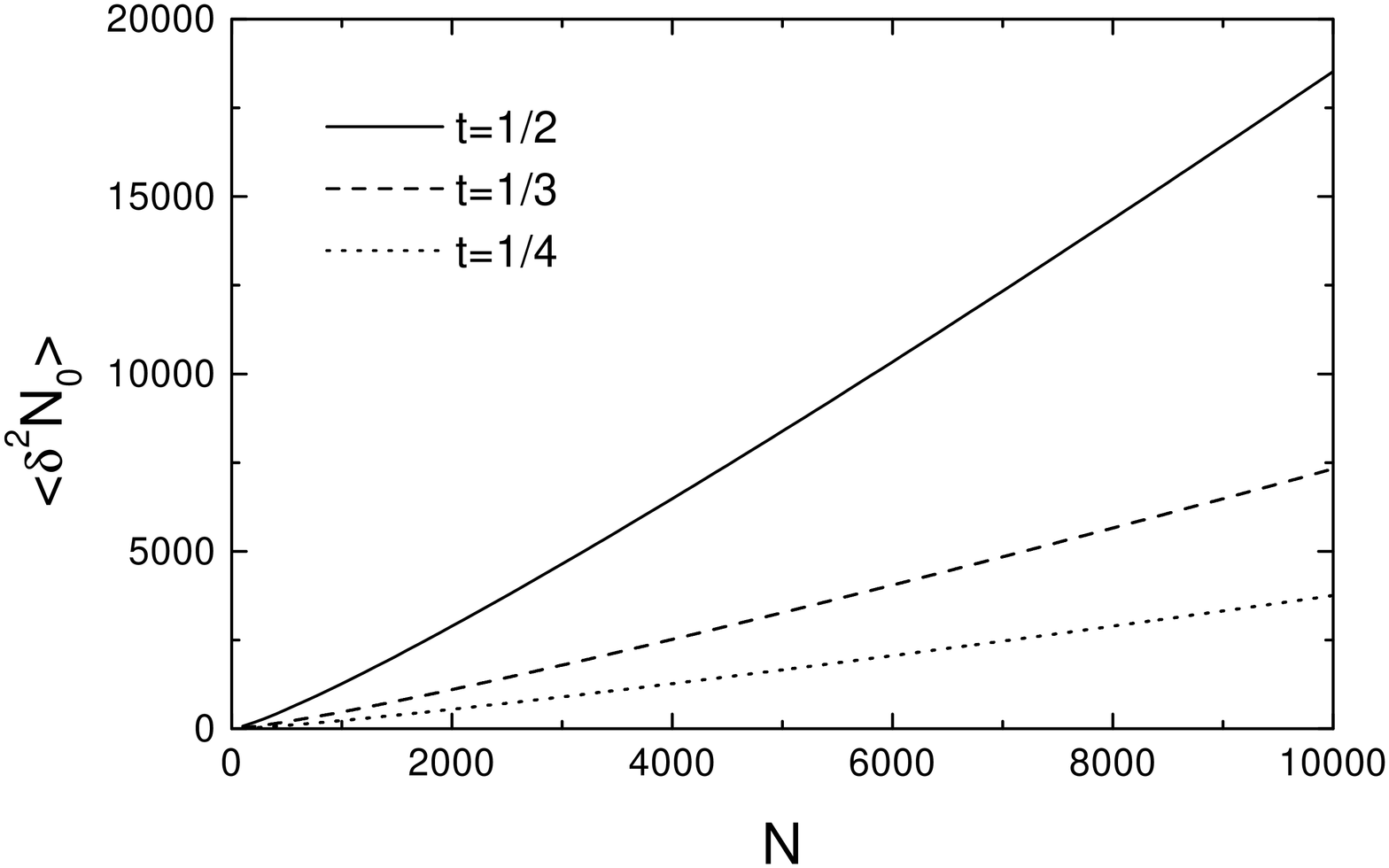,width=\columnwidth,angle=270}
\caption{Displayed is ${\left\langle \delta ^{2}N_{{\bf 0}}\right\rangle }$ as
a function of $N$ bosons in a 2D harmonic trap for various reduced
temperatures. The linearity of ${\left\langle \delta ^{2}N_{{\bf 0}%
}\right\rangle }$ about $N$ is clearly shown in the figure. This shows that
the behavior of ${\left\langle \delta ^{2}N_{{\bf 0}}\right\rangle }$ can be
approximated as normal for 2D harmonically trapped Bose gas.
}
\end{figure}

We see that the particle number fluctuations of the 2D
harmonically trapped Bose gas exhibit a very weak anomalous
behavior,  as it is controlled by the factor $\ln \left( Nt^{2}/\zeta
\left( 2\right) \right) $. In Fig. 2 we plot the result of
${\left\langle \delta ^{2}N_{%
{\bf 0}}\right\rangle }$ as a function of $N$ for various reduced
temperatures. The linear property of ${\left\langle \delta ^{2}N_{{\bf 0}%
}\right\rangle }$ about $N$ is clearly illustrated. This shows
that the behavior of ${\left\langle \delta ^{2}N_{{\bf
0}}\right\rangle }$ can be approximated as normal.

Shown in Fig. 3 is the result for $\left\langle N_{{\bf {0}}}\right\rangle /N$ as
a function of temperature for $N=10^{3}$ 2D harmonically trapped
bosons. The solid line displays $\left\langle N_{{\bf
{0}}}\right\rangle /N$ using a grand-canonical ensemble (or
$N_{{\bf {0}}}^{p}$ within the canonical ensemble). The dotted
line displays $\left\langle N_{{\bf {0}}}\right\rangle /N$ within
the canonical ensemble. In Fig. 4 we plot the numerical result of
$\delta N_{{\bf 0}}$ for $N=10^{3}$ trapped 2D ideal bosons. From
Fig. 4 we find that the particle number fluctuations of the 2D
Bose gas are larger than those obtained for the trapped 3D Bose gas
at identical reduced temperature (see Fig. 1).

\begin{figure}[tb]
\psfig{figure=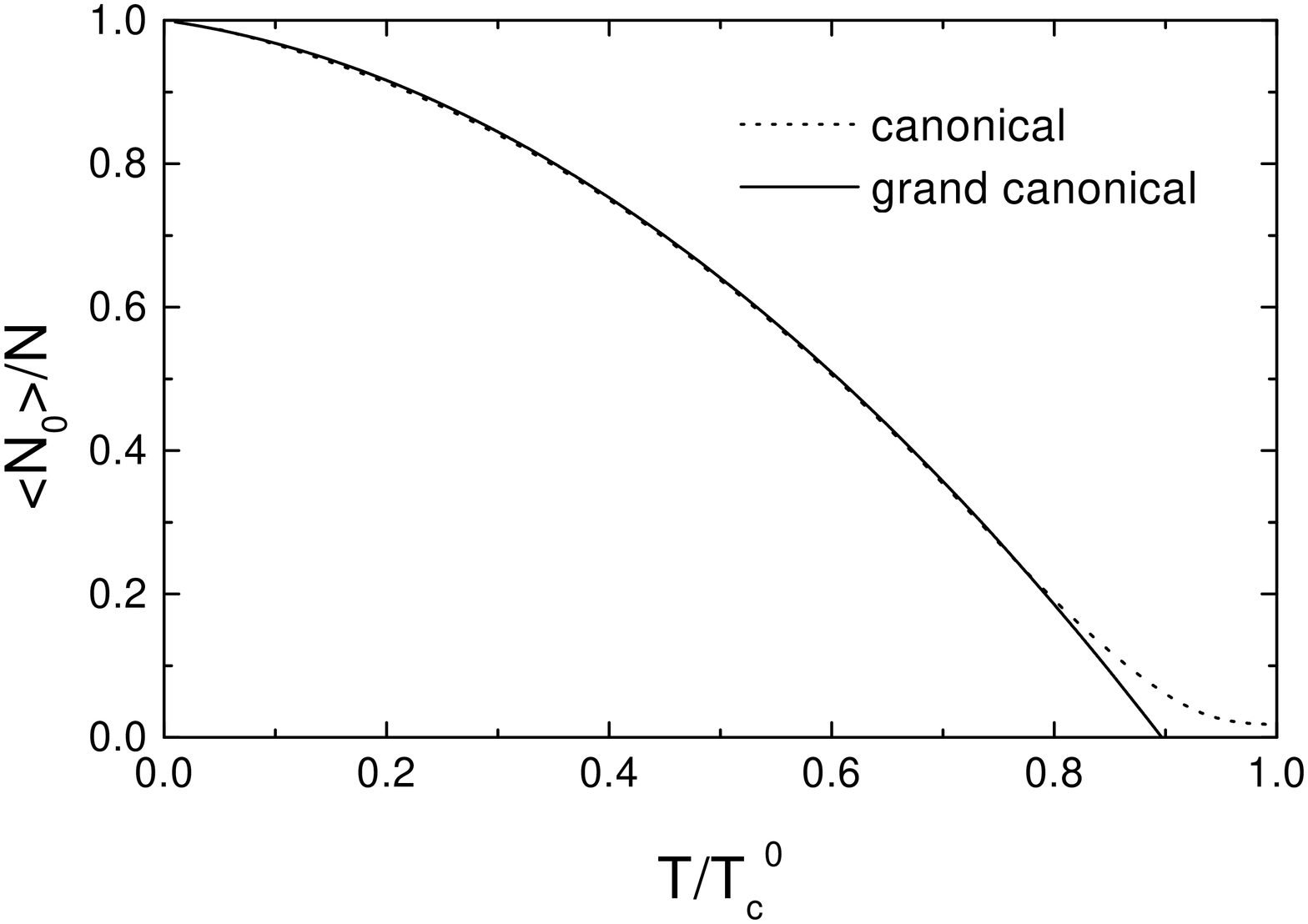,width=\columnwidth,angle=270}
\caption{Relative mean ground state occupation number ${\left\langle N_{{\bf 0}
}\right\rangle /N}$ vs $T/T_{c}^{0}$ for $N=10^{3}$ non-interacting bosons
confined in a 2D harmonic trap. The solid line shows ${\left\langle N_{{\bf 0
}}\right\rangle /N}$ within the grand-canonical ensemble, while the dotted
line displays ${\left\langle N_{{\bf 0}}\right\rangle /N}$ within the
canonical ensemble.
}
\end{figure}

\begin{figure}[tb]
\psfig{figure=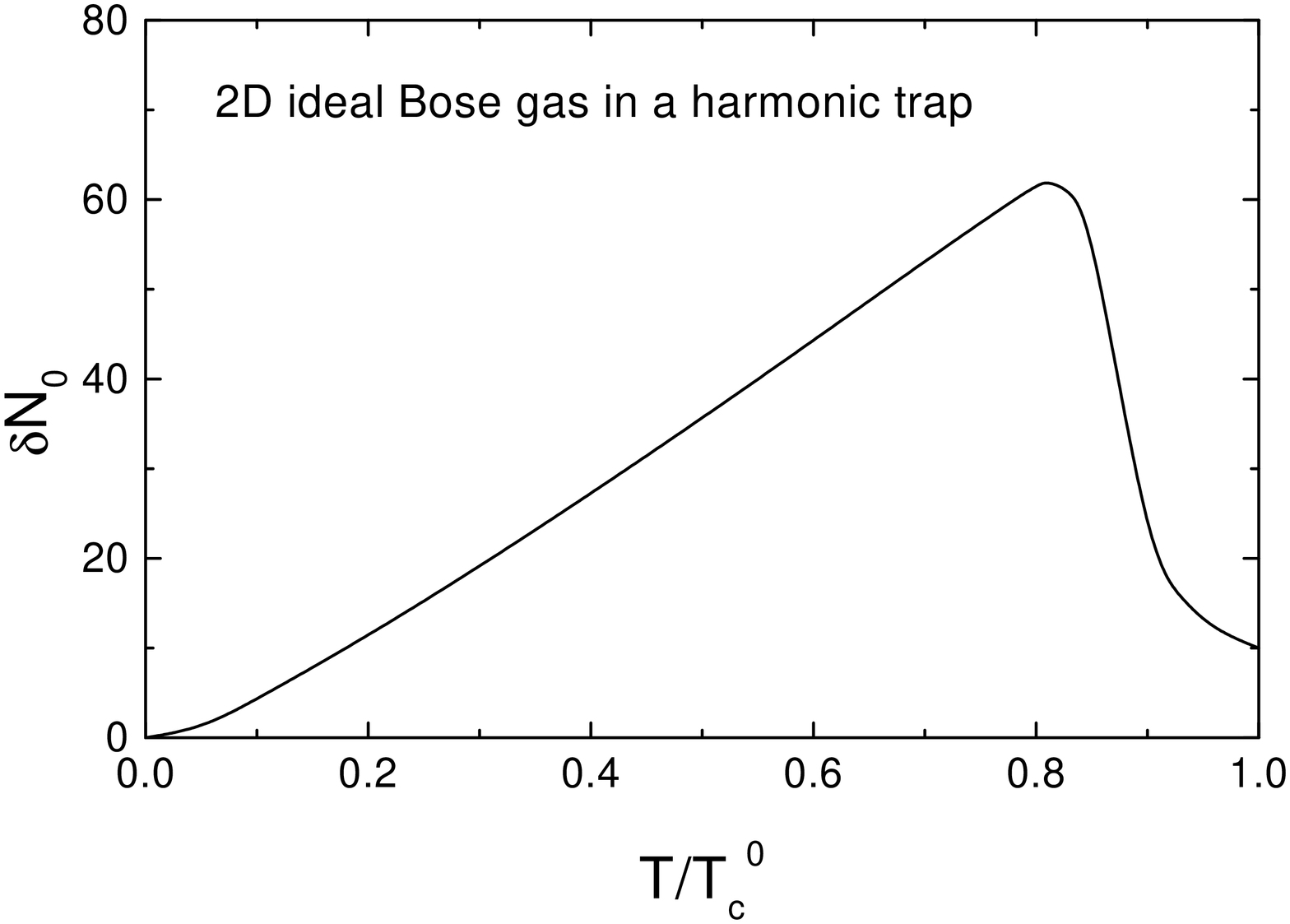,width=\columnwidth,angle=270}
\caption{Displayed is $\delta N_{{\bf {0}}}$ for $N=10^{3}$ non-interacting
bosons confined in a 2D harmonic trap.
}
\end{figure}

For a 2D ideal Bose gas confined in a box, the single-particle
energy level is given by ${{\varepsilon _{{\bf {n}}}=\pi ^{2}\hbar
^{2}\left( n_{x}^{2}+n_{y}^{2}\right) /2mL^{2}}}$. There exists a
logarithmic divergence in Eqs. (\ref{main3}) and (\ref{non1}) when
the sum is calculated by integration in momentum space. The
occurrence of the logarithmic divergence implies an anomalous
behavior of the particle number
fluctuations. Because of the divergence in Eqs. (\ref{main3}) and (\ref{non1}%
), the sum can not be calculated by integration, but is dominated by the
discretized sum over the low-energy bosons. The finial result of the
particle number fluctuations can be estimated as

\begin{equation}
{\left\langle \delta ^{2}N_{{\bf 0}}\right\rangle \approx \sum_{{\bf n}\neq
{\bf 0}}}\frac{1}{\left( n_{x}^{2}+n_{y}^{2}\right) ^{2}}\left( \frac{%
2mL^{2}k_{B}T}{\pi ^{2}\hbar ^{2}}\right) ^{2}{.}  \label{2D-flu-box}
\end{equation}

\noindent This result given shows that, for the 2D gas confined in  box,  there is a
strong anomalous behavior
for  the particle number fluctuations, ${\it i.e.}$, ${\left\langle \delta 
^{2}N_{{\bf 0}%
}\right\rangle \sim N}^{2}$.


\section{One-Dimensional Bose Gases}


The single-particle energy level of a 1D harmonically confined
ideal Bose gas has the form

\begin{equation}
{\varepsilon _{n}=\left( n+\frac{1}{2}\right) \hbar \omega _{1D}.}
\label{1D-energy}
\end{equation}

\noindent From Eq. (\ref{main3}) one obtains

\begin{equation}
{N_{0}=N-\sum_{n=1}^{\infty }\frac{1}{\exp \left[ \beta n\hbar \omega
_{1D}\right] \exp \lbrack -{\alpha \left( N,N_{{\bf {0}}}\right) }\rbrack -1}%
.}  \label{1DN}
\end{equation}

\noindent The most probable value $N_{0}^{p}$ reads

\begin{equation}
{N_{0}^{p}=N-\sum_{n=1}^{\infty }\frac{1}{\exp \left[ \beta n\hbar \omega
_{1D}\right] -1}.}  \label{1DNp}
\end{equation}

\noindent There is a logarithmic divergence when the sum is calculated by integration
in momentum space. With the method developed in Ref. \cite{KETTERLE}, the
approximation of ${N_{0}^{p}}$ is given by

\begin{equation}
{N_{0}^{p}=N-}\left( \frac{k_{B}T}{\hbar \omega _{1D}}\right) \ln \frac{%
2k_{B}T}{\hbar \omega _{1D}}{.}  \label{1DNp1}
\end{equation}

\noindent The critical temperature $T_{1D}$ of the trapped 1D Bose gas can be obtained by setting 
${N_{0}^{p}=0}$ in Eq. (\ref{1DNp1}). We obtain
$T_{1D}=(\hbar\omega_{1D}/k_{B}) (N/
ProductLog\left[ 2N\right])$, where $ProductLog\left[ z\right] $
 is the solution $w=ProductLog\left[ z\right] $
of the equation $z=w\,e^{w}$ (see Ref.\cite{MATHEMICS}).

From Eqs. (\ref{1DN}) and (\ref{1DNp}) we get

$${
{N}_{0}{-N_{0}^{p}}=\sum_{n=1}^{\infty }
\left[
\frac{1}
{\exp \left[ \beta
n\hbar \omega _{1D}\right] -1}
\right.
}$$

\begin{equation}
~~~~~~~~~~~~~\left. 
-\frac{1}
{
\exp \left[ \beta n\hbar \omega_{1D}\right] \exp \lbrack -{\alpha \left( N,N_{0}\right) }\rbrack -1
}
\right]~.  \label{N0Np}
\end{equation}

\noindent The occurrence of the logarithmic divergence in momentum space
implies that the leading contribution to the particle number
fluctuations comes from the low energy bosons. In this situation, Eq. (\ref{N0Np})
can be estimated to be

\begin{equation}
{N_{0}-N_{0}^{p}=\sum_{n=1}^{\infty }}\left[ {\frac{1}{\beta n\hbar \omega
_{1D}}-\frac{1}{\beta n\hbar \omega _{1D}-{\alpha \left( N,N_{0}\right) }}}%
\right] {.}
\end{equation}

Assuming

\begin{equation}
\left| {\alpha \left( N,N_{0}\right) }\right| <<\beta \hbar \omega _{1D}{,}
\label{1Dapp}
\end{equation}

\noindent we obtain the final result of ${\alpha \left( N,N_{0}\right) }$:

\begin{equation}
{\alpha \left( N,N_{0}\right) =}-\frac{N_{0}-N_{0}^{p}}{\sum_{n=1}^{\infty
}1/\lbrack \beta n\hbar \omega _{1D}\rbrack ^{2}}{.}  \label{1D-alpha}
\end{equation}

\noindent It is easy to find that the result given by Eq.
(\ref {1D-alpha}) coincides with the assumption given by Eq.
(\ref{1Dapp}). The probability distribution of the  1D
harmonically trapped Bose gas is then given by

\begin{equation}
G_{1D}{\left( N,N_{0}\right) =A}_{1D}{\exp }\left[ -\frac{\left(
N_{0}-N_{0}^{p}\right) ^{2}}{2\sum_{n=1}^{\infty }1/\lbrack \beta n\hbar
\omega _{1D}\rbrack ^{2}}\right] {,}  \label{1D-dis}
\end{equation}

\noindent where $A_{1D}$ is a normalization constant.

Using the probability distribution $G_{1D}{\left( N,N_{0}\right) }$, one
obtains the analytical result of ${\left\langle \delta
^{2}N_{0}\right\rangle }$ below the critical temperature:

\begin{equation}
{\left\langle \delta ^{2}N_{0}\right\rangle =\zeta }\left( 2\right) \left[
\frac{Nt}{ProductLog\left[ 2N\right] }\right] ^{2}{.}  \label{1D-flu}
\end{equation}

\noindent At $T_{1D}$, ${\left\langle \delta ^{2}N_{0}\right\rangle }$ can be
approximated as

\begin{equation}
{\left\langle \delta ^{2}N_{0}\right\rangle |}_{T=T_{1D}}{=}\left( 1-\frac{2%
}{\pi }\right) {\zeta }\left( 2\right) \left[ \frac{N}{ProductLog\left[
2N\right] }\right] ^{2}{.}  \label{1D-near}
\end{equation}

Comparing this result with  those obtained for trapped
2D and 3D Bose gases, the {particle number fluctuations in  1D
have a much stronger dependence on }$N$. This
fact reminds us of the occurrence of infrared divergence arising
from the low energy bosons in a 1D free ideal Bose gas. Although
here we consider a confined Bose gas and hence  the infrared
divergence has been cut off in the discrete sums in Eqs.
(\ref{1DN}) and (\ref{1DNp}), the existence of low energy bosons
results in an anomalous behavior of the particle number
fluctuations. From Eqs. (\ref{1D-flu}) and (\ref{1D-near}), it is
obvious that the particle number fluctuations of the trapped 1D
ideal Bose gas exhibit a very strong anomalous behavior. In Fig. 5 we plot ${%
\left\langle \delta ^{2}N_{{\bf 0}}\right\rangle }$ as a function
of $N$ for various reduced temperatures. The anomalous behavior of
${\left\langle \delta ^{2}N_{{\bf 0}}\right\rangle }$ for the 1D
harmonically trapped Bose gas is clearly illustrated in the figure.

\begin{figure}[tb]
\psfig{figure=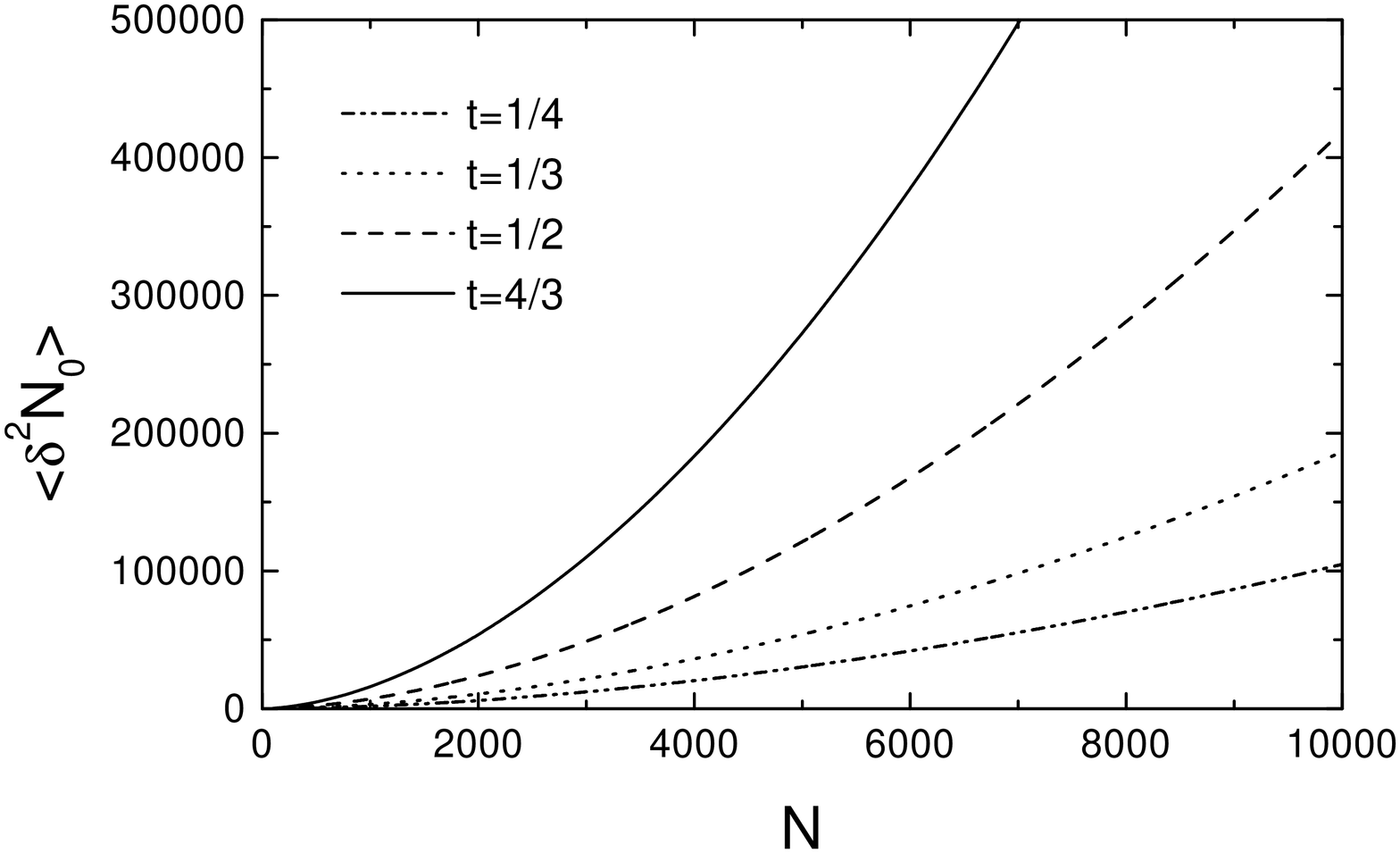,width=\columnwidth,angle=270}
\caption{We plot ${\left\langle \delta ^{2}N_{{\bf 0}}\right\rangle }$ as a
function of $N$ bosons in a 1D harmonic trap for various reduced
temperatures. The anomalous behavior of ${\left\langle \delta ^{2}N_{{\bf 0}%
}\right\rangle }$ for 1D harmonically trapped Bose gas is clearly
demonstrated in the figure.}
\end{figure}

Shown in Fig. 6 is the numerical result of $\left\langle N_{{\bf 0}%
}\right\rangle /N$ for $N=10^{3}$ trapped 1D ideal bosons. The numerical result of 
$\delta N_{{\bf 0}}$ for $N=10^{3}$  is plotted in Fig. 7. At
finite temperature, this sort of large condensate thermal
fluctuations may destabilize the condensate. The inclusion of
repulsive interactions between atoms can, however, significantly
change the behavior of particle number fluctuations. An
interacting Bose gas exhibits a phonon-type low energy
excitations. The fluctuations may still exhibit an  anomalous
behavior even for the 1D interacting Bose gas confined in a
harmonic trap. We expect that the two-body repulsive force may
lower the particle number fluctuations significantly and lead to
the stability of 1D trapped condensate.

\begin{figure}[tb]
\psfig{figure=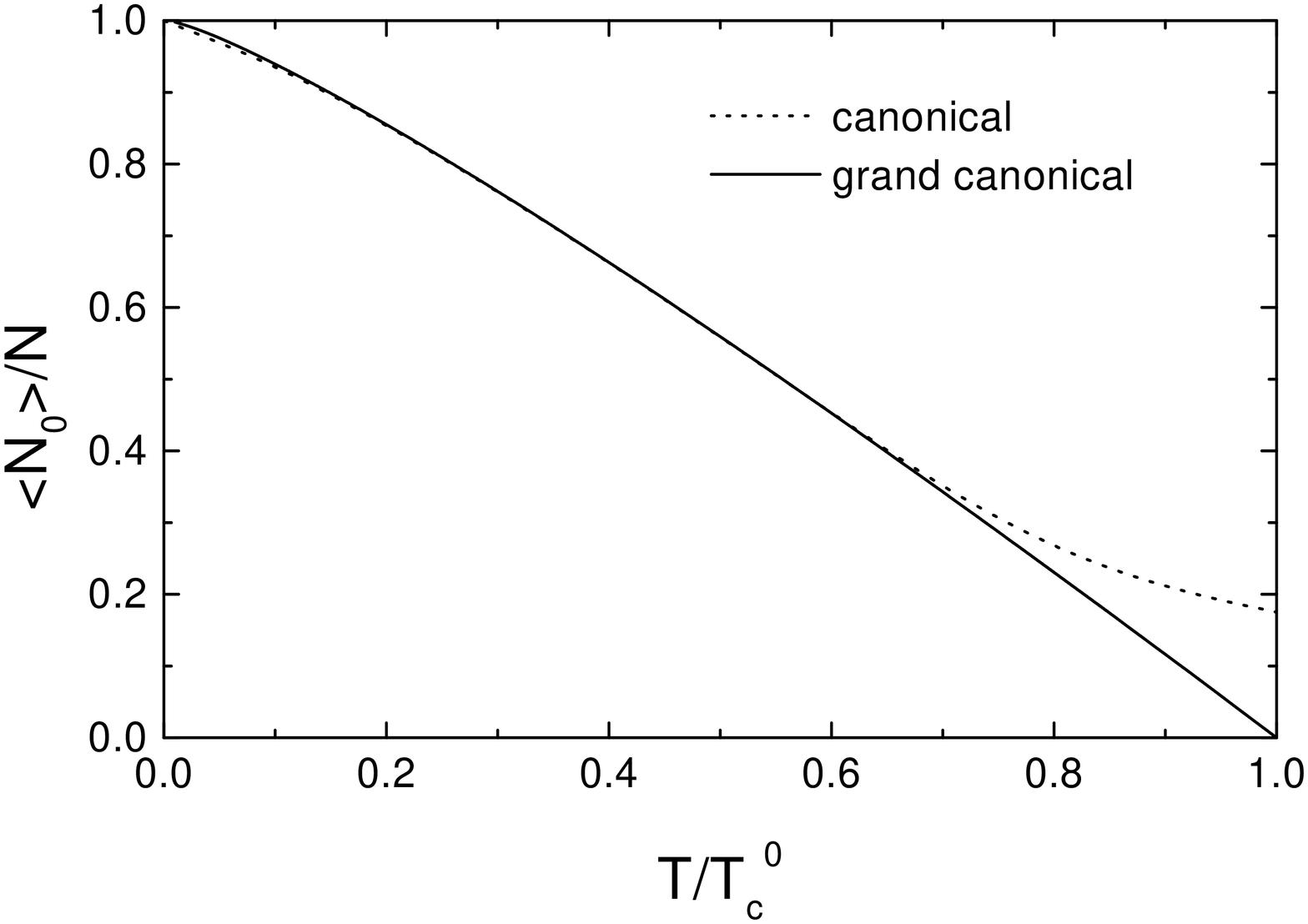,width=\columnwidth,angle=270}
\caption{Displayed is ${\left\langle N_{{\bf 0}}\right\rangle /N}$ vs $%
T/T_{c}^{0}$ for $N=10^{3}$ non-interacting bosons confined in a 1D harmonic
trap. The solid line shows ${\left\langle N_{{\bf 0}}\right\rangle /N}$
within the grand-canonical ensemble, while the dotted line displays ${%
\left\langle N_{{\bf 0}}\right\rangle /N}$ within the canonical ensemble.}
\end{figure}

\begin{figure}[tb]
\psfig{figure=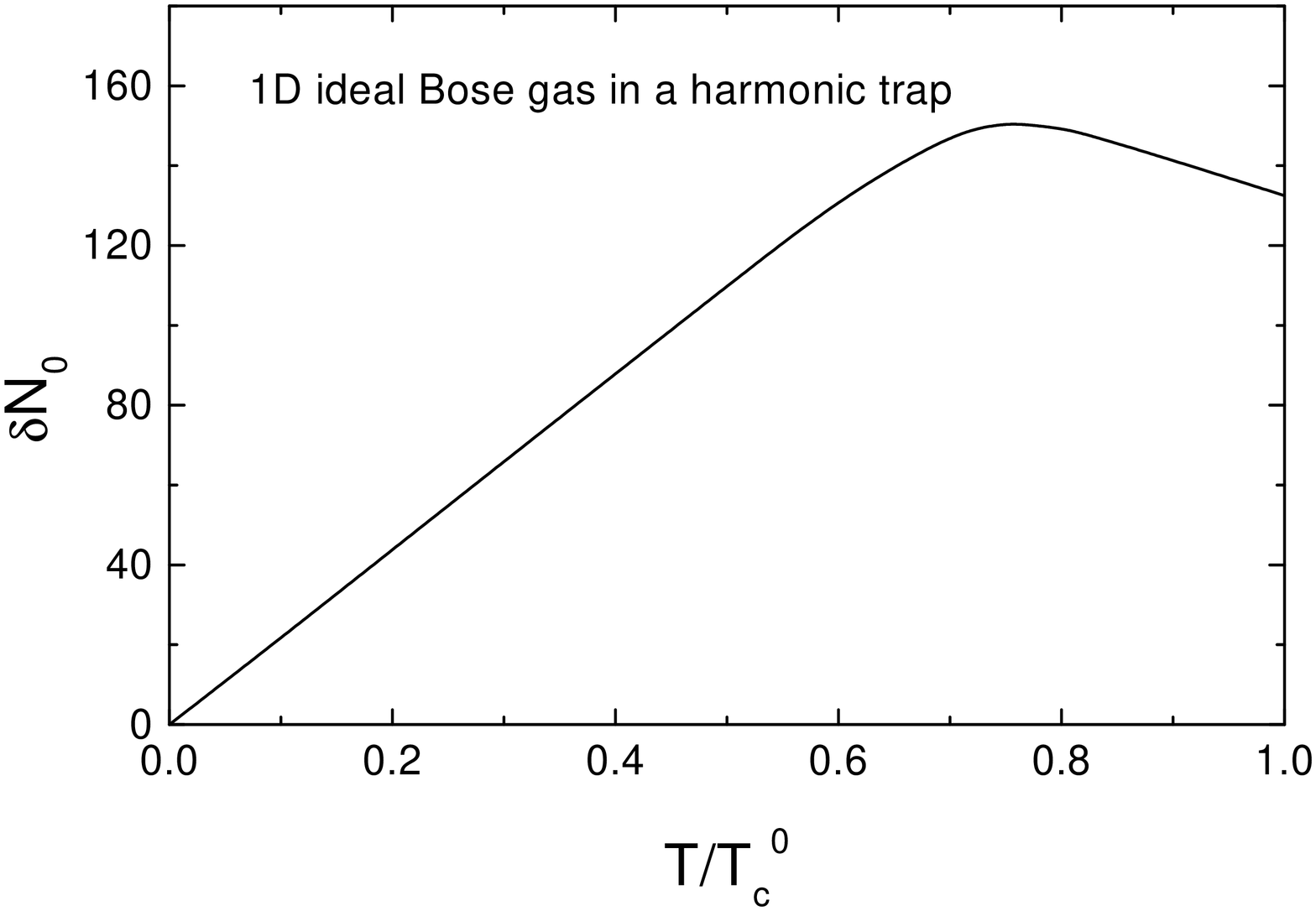,width=\columnwidth,angle=270}
\caption{The numerical result of $\delta N_{{\bf {0}}}$ for $N=10^{3}$
non-interacting bosons confined in a 1D harmonic trap.}
\end{figure}

For an ideal Bose gas confined in a  line (1D ``box''), the  particle number
fluctuations exhibit a much stronger anomalous behavior in comparison
with the case of the
harmonically trapped Bose gas. This can be seen because there
is a $1/n$ divergence when $n\rightarrow 0$ in Eqs. (\ref{main3}) and (\ref{non1}). 
The finial result of the particle number fluctuations in this case is given
by

\begin{equation}
{\left\langle \delta ^{2}N_{{0}}\right\rangle \approx }\left( {\sum_{%
{n=1}}^{\infty }}\frac{1}{n^{4}}\right) \left( \frac{2mL^{2}k_{B}T}{\pi
^{2}\hbar ^{2}}\right) ^{2}{.}  \label{1D-flu-box}
\end{equation}

\noindent Thus in such system there is a very strong
anomalous behavior
for the particle number fluctuations, {\it i.e.}, ${\left\langle \delta ^{2}N_{{ 
0}}\right\rangle \sim N}^{2}$. This sort of anomalous behavior will
certainly destabilize the condensate.


\section{Discussion and Conclusion}


In this work, a saddle-point method has been developed to
investigate the canonical partition function of trapped Bose
gases. Within the canonical ensemble the analytical probability
distribution and particle number fluctuations have been obtained
for various dimensions, especially in one and two dimensions.
Different from the conventional methods, here the analytical
probability distribution is obtained directly from the canonical
partition function of the system. Using the probability
distribution function, we have calculated the thermodynamic
properties of the trapped Bose gas, such as condensate fraction
and particle number fluctuations. Through the calculations of the
probability distribution function, we have provided a simple
method recovering the applicability of the saddle-point method for
studying the particle number fluctuations.

We have found that although the probability distribution of the
harmonically trapped Bose gas are Gaussian functions, the
behavior of the particle number fluctuations is quite different
from each other for  different dimensions. 
For a trapped 2D ideal Bose gas,  the fluctuations
exhibit a very weak anomalous behavior, while there is a strong
anomalous behavior for a trapped 1D ideal Bose gas.
These properties are clearly shown by the explicit formulas given by
(\ref{analytical}), (\ref{2D-flu}) and (\ref{1D-flu}) for  the particle number 
fluctuations in
three, two and one dimensions, respectively.
We expect that the recent
realization of BEC in lower dimensions\cite{EXPLOW} makes it very
promising to explore the particle number fluctuations of
lower-dimensional condensates.
Our results also show that different  confinements
can significantly change the effect of the particle number
fluctuations.  Comparing with the case  confined in a harmonic
trap, the Bose gases confined in a box exhibit much stronger
effect for the condensate fluctuations.

It is obvious that the method developed here can be applied to
other Bose systems. Based on our approach, one can also formulate
an improved saddle-point method to calculate the change of the
particle number fluctuations when  using  a microcanonical
ensemble. It can be  shown that this will result in a change
for particle number fluctuations
only  with a numerical prefactor \cite {GAJ,GRO,NAVEZ,IDZ}.


\section*{Acknowledgments}


This work was supported by the Science Foundation of Zhijiang
College, Zhejiang University of Technology and Natural Science
Foundation of Zhejiang Province, China. G. X. Huang was supported
by the National Natural Science Foundation of China, the
Trans-Century Training Programme Foundation for the Talents and
the University Key Teacher Foundation of Chinese Ministry of
Education. S. J. Liu and H. W. Xiong thank Professors G. S. Jia
and J. F. Shen for their enormous encouragement.


\section*{Appendix}


The probability distribution of a harmonically trapped Bose gas is
a Gaussian function for various dimensions. The ratio between $P({N_{{\bf 0}}|N%
})$ and $P({N_{{\bf 0}}^{p}|N})$ is given by

\begin{equation}
{G}\left( {N_{{\bf 0}},N_{{\bf 0}}^{p}}\right) {=}\exp \left[ -a\left( {N_{%
{\bf 0}}-N_{{\bf 0}}^{p}}\right) ^{2}\right] {,}  \label{a1}
\end{equation}

\noindent where $a$ is a constant and is related to the confinement,
dimensionality and the total number of particles of the system. The sum in
Eq. (\ref{fluc-ideal}) can be replaced by an integral

$${
\left\langle \delta ^{2}N_{{\bf {0}}}\right\rangle =\frac{\int_{N_{{\bf {0}}
}=0}^{N}N_{{\bf {0}}}^{2}G\left( N,N_{{\bf {0}}}\right) dN_{{\bf {0}}}}{
\int_{N_{{\bf {0}}}=0}^{N}G\left( N,N_{{\bf {0}}}\right) dN_{{\bf {0}}}}
}$$

\begin{equation}
{~~~~~~~~~~~~~~~~~~~~~~-\left[ \frac{\int_{N_{{\bf {0}}}=0}^{N}N_{{\bf {0}}}G\left( N,N_{{\bf {0}}
}\right) dN_{{\bf {0}}}}{\int_{N_{{\bf {0}}}=0}^{N}G\left( N,N_{{\bf {0}}
}\right) dN_{{\bf {0}}}}\right] ^{2}.}  \label{a2}
\end{equation}

Using the integral transformation $x={N_{{\bf 0}}-N_{{\bf 0}}^{p}%
}$, one obtains

\begin{equation}
\int_{N_{{\bf {0}}}=0}^{N}N_{{\bf {0}}}^{2}G\left( N,N_{{\bf {0}}}\right)
dN_{{\bf {0}}}=\int_{-N_{{\bf {0}}}^{p}}^{N-N_{{\bf {0}}}^{p}}(x+N_{{\bf {0}}%
}^{p})^{2}\exp \lbrack -ax^{2}\rbrack dx{.}  \label{a3}
\end{equation}

\noindent For the temperature below the critical temperature, $N_{{\bf {0}}}^{p}>>1$.
Hence the integral in Eq. (\ref{a3}) can be estimated to be

\begin{equation}
\int_{N_{{\bf {0}}}=0}^{N}N_{{\bf {0}}}^{2}G\left( N,N_{{\bf {0}}}\right)
dN_{{\bf {0}}}\approx \int_{-\infty }^{\infty }(x+N_{{\bf {0}}}^{p})^{2}\exp
\lbrack -ax^{2}\rbrack dx{.}  \label{a4}
\end{equation}

\noindent Similarly, we have

\begin{equation}
\int_{N_{{\bf {0}}}=0}^{N}G\left( N,N_{{\bf {0}}}\right) dN_{{\bf {0}}%
}\approx \int_{-\infty }^{\infty }\exp \lbrack -ax^{2}\rbrack dx{,}
\label{a5}
\end{equation}

\noindent and

\begin{equation}
\int_{N_{{\bf {0}}}=0}^{N}N_{{\bf {0}}}G\left( N,N_{{\bf {0}}}\right) dN_{%
{\bf {0}}}\approx \int_{-\infty }^{\infty }(x+N_{{\bf {0}}}^{p})\exp \lbrack
-ax^{2}\rbrack dx{.}  \label{a6}
\end{equation}

Using the formulas (\ref{a4}), (\ref{a5}) and (\ref{a6}), it is
straightforward to obtain the analytical result of ${\left\langle \delta
^{2}N_{{\bf {0}}}\right\rangle }$ for the temperature below the critical
temperature

\begin{equation}
{\left\langle \delta ^{2}N_{{\bf {0}}}\right\rangle =}\frac{1}{2a}{.}
\label{a7}
\end{equation}

\noindent We can see from Eq. (\ref{a7}) that the behavior of the fluctuations is
determined by the factor $a$.

At the critical temperature in the thermodynamic limit, we
have $N_{{\bf {0}}}^{p}=0$. The probability distribution of the
system is then

\begin{equation}
{G}\left( {N_{{\bf 0}},N_{{\bf 0}}^{p}}\right) {=}\exp \left[ -a{N_{{\bf 0}%
}^{2}}\right] {.}  \label{a8}
\end{equation}

After a simple calculation, the analytical result of ${\left\langle \delta
^{2}N_{{\bf {0}}}\right\rangle }$ at the critical temperature is given by

\begin{equation}
{\left\langle \delta ^{2}N_{{\bf {0}}}\right\rangle =}\left( 1-\frac{2}{\pi }%
\right) \frac{1}{2a}{.}  \label{a9}
\end{equation}


\begin{references}
\bibitem{MATHEMICS}  In software Mathematic 4.0,$\ ProductLog\left[ z\right]
$ is defined as the solution for $w$ in $z=we^{w}$, and is easily dealed
with.
\end{references}
\end{document}